\documentstyle[11pt]{article}

\begin{document}
\def\beq{\begin{equation}}
\def\eeq{\end{equation}}
\def\bsigma{\mbox{\boldmath $\sigma$}}
\def\balpha{\mbox{\boldmath $\alpha$}}

\makeatletter
\renewcommand\theequation{\thesection.\arabic{equation}}
\@addtoreset{equation}{section}
\makeatother

\baselineskip 20pt
\rightline{CU-TP-882}
\rightline{CLNS-98/1548}
\rightline{hep-th/9803164}
\vskip 1cm
\centerline{\Large\bf Explicit Multimonopole Solutions}
\centerline{\Large\bf in SU(N) Gauge Theory}
\vskip 1cm
\centerline{\large\it
Erick J. Weinberg $^a$\footnote{electronic mail: ejw@phys.columbia.edu}  
and Piljin Yi $^b$\footnote{electronic mail: piljin@mail.lns.cornell.edu}}
\vskip 1mm
\centerline{$^a$Physics Department, Columbia University, New York, NY 10027}
\vskip 1mm
\centerline{$^b$F.R. Newman Laboratory of Nuclear Studies, Cornell
University, Ithaca, NY 14853}

\vskip 1.5cm
\centerline{\bf ABSTRACT}
\vskip 0.5cm
\begin{quote}

We construct multimonopole solutions containing $N-1$ distinct
fundamental monopoles in $SU(N)$ gauge theory.  When the gauge
symmetry is spontaneously broken to $U(1)^{N-1}$, the monopoles are
all massive, and we show that the fields can be written in terms of
elementary function for all values of the monopole positions and
phases.  In the limit of unbroken $U(1)\times SU(N-2) \times U(1)$
symmetry, the configuration can be viewed as containing a pair of
massive monopoles, each carrying both $U(1)$ and $SU(N-2)$ magnetic
charges, together with $N-3$ massless monopoles that condense into a
cloud of non-Abelian fields.  We obtain explicit expressions for the
fields in the latter case and use these to analyze the properties of
the non-Abelian cloud.

\end{quote}

\newpage
\setcounter{footnote}{0}

\section{Introduction}

The massive magnetic monopoles of spontaneously broken gauge theories have long
been the objects of considerable study.  Although these monopoles arise as
spatially extended solutions of the classical field equations, they correspond
to single-particle states of the full quantum theory.  Their dynamics, at least
at low energies, can be described by a small number of degrees of freedom, just
like that of the elementary particles of the theory \cite{manton}. Indeed, 
these magnetically charged states can be regarded as the counterparts, 
related by an exact duality symmetry in certain supersymmetric theories, 
of the  massive electrically charged states built from the elementary 
excitations of the theory \cite{olive}.
 
Recent studies of low-energy monopole dynamics have shown that when
the unbroken gauge group has a non-Abelian component there are degrees
of freedom that can be attributed to the presence of massless
non-Abelian monopoles; these can be seen as the dual counterparts of
the massless elementary gauge bosons \cite{nonabelian}.  In contrast
with the massive monopoles, these massless monopoles cannot be
exhibited as isolated classical solutions, but can be studied
classically only as part of multimonopole configurations.  In the
simplest example, an $SO(5)$ solution \cite{so5} with one massive and
one massless monopole, the massless monopole is manifested as a
spherically symmetric ``cloud'' of non-Abelian fields surrounding the
massive monopole.  In this paper we examine a somewhat more complex
class of configurations containing two massive and $N-3$ massless
monopoles in the Bogomolny-Prasad-Sommerfield (BPS) \cite{bps} limit
of an $SU(N)$ gauge theory.  By obtaining explicit analytic
expressions for the gauge and Higgs fields we can see clearly how the
massless monopoles condense into a non-Abelian cloud and can verify
that the properties of this cloud inferred from the form of the moduli
space metric are indeed present.
 
The origin of these massless monopoles can be understood by recalling
that the magnetically charged BPS solutions for an arbitrary gauge
group $G$ can all be analyzed in terms of fundamental monopoles of
various types.  The simplest case is when the adjoint Higgs field
breaks a group of rank $r$ maximally, to $U(1)^r$.  There are then $r$
quantized topological charges, one for each of the unbroken $U(1)$
factors.  Corresponding to each of these is a fundamental monopole
solution carrying a single unit of topological charge.  Each 
of these is described by four collective coordinates, three specifying its
position and one corresponding to an overall $U(1)$ phase.  These
solutions can be realized explicitly by embedding the unit $SU(2)$
monopole using a preferred set of simple roots.  All higher charged
solutions may be regarded as multimonopole solutions containing
appropriate numbers of the various fundamental monopoles.  Not only
does one find that the energy of the solution is the sum of the masses
of the component monopoles, but an index theory analysis shows that
the number of zero modes, and hence of collective coordinates, is
precisely four times the number of component monopoles \cite{erick}.
 
This can be illustrated in a particularly simple fashion when the
gauge group is $SU(N)$.  By means of a gauge transformation, the
asymptotic value of the Higgs field in any fixed direction can always
be brought into the form
\begin{equation}
   \Phi = {\rm diag}\,\, (t_N, t_{N-1}, \dots, t_1)
\label{phievalues}
\end{equation}
with $t_1 \le t_2 \le \dots \le t_N$.  We may write the asymptotic magnetic
field in the same direction as  $F_{ij} = \epsilon_{ijk} r_k Q_M/r^3$, with 
\begin{equation}
   Q_M ={4\pi\over e} {\rm diag}\,\, (n_{N_1}, n_{N-1}-n_{N-2}, \dots,
           n_2-n_1, -n_1 )  \, .
\label{Qevalues}   
\end{equation}  
The generalized topological quantization condition implies that the $n_j$
must be integers \cite{gno}.    

If the eigenvalues of the asymptotic Higgs field are all distinct, then the
unbroken gauge group is $U(1)^r$ and the $n_j$ are the topological charges. 
Apart from a constant Higgs field contribution, the $k$th fundamental monopole,
with $n_j=\delta_{jk}$, is obtained by embedding the $SU(2)$ monopole
solution (rescaled appropriately so as to give the correct Higgs
expectation value) in the $2\times 2$ block at the intersections of the
$(N-k)$th  and $(N+1-k)$th rows and columns.  The resulting $SU(N)$ monopole
then has a mass \begin{equation}
     M_k = {2\pi(t_{k+1} - t_k) \over e}
\label{massformula}
\end{equation}    
where $e$ is the gauge coupling.  Although there are other possible
$SU(2)$ embeddings, both the mass formulas and the zero mode counting
indicate that these are merely special cases of multimonopole
solutions.   

Varying the asymptotic Higgs field so that two or more of its
eigenvalues are equal enlarges the unbroken symmetry group so that
some of the $U(1)$ factors are replaced by a non-Abelian group
$K$. The magnetic charge must still be of the form of
Eq.~(\ref{Qevalues}), but the $n_j$ that correspond to roots of $K$
are no longer topologically conserved charges; in fact, they are not
even gauge-invariant.  According to Eq.~(\ref{massformula}), the
corresponding monopoles should become massless in this limit.  From
the point of view of electric-magnetic duality this seems quite
reasonable, since one would expect the massless gauge bosons carrying
electric-type charges in the subgroup $K$ to have massless
counterparts carrying magnetic charges.  On the other hand, one would
not expect to find zero energy solitons.  Indeed, the classical
one-monopole solution tends toward the vacuum solution as the limit of
unbroken symmetry is approached.  However,
examination \cite{nonabelian} of the moduli space Lagrangian that
describes the low-energy dynamics of a collection of BPS monopoles
suggests that the degrees of freedom corresponding to these monopoles
can survive even in the massless limit.  Specifically, if a number of
massless monopoles are combined with one or more massive monopoles to
give an $n$-monopole solution whose total magnetic charge is
invariant\footnote{One encounters a number of pathologies
when dealing with configurations whose total magnetic charge has a non-Abelian
component \cite{color}.  Requiring that the total charge be purely Abelian is
not really a significant restriction, since the additional monopoles needed to
meet this requirement can be placed arbitrarily far from the original ones.} 
under $K$, the dimension of the moduli space, and hence the number of collective
coordinates, remains $4n$ even in the limit of non-Abelian unbroken symmetry
\cite{weinberg}. Furthermore, examination of specific examples suggests that the
moduli space metric, and hence the Lagrangian, behaves smoothly in this
limit.

As noted above, the simplest examples, containing one massive and one massless
monopole, arise in the context of an $SO(5)$ gauge theory spontaneously broken
to $SU(2)\times U(1)$ \cite{so5}.  These contain a massive monopole core 
surrounded by a
spherically symmetric ``non-Abelian cloud'', of arbitrary radius, that can be
viewed as the remnant of the massless monopole.  Within the cloud there is a
Coulomb magnetic field corresponding to a magnetic charge with components lying
both in the unbroken $U(1)$ and in the unbroken $SU(2)$, while outside the
cloud only the $U(1)$ Coulomb field is present.  The solution is described by
eight collective coordinates.  Four of these are readily identified as the
position and $U(1)$ phase of the massive monopole.   The other four coordinates
describe the cloud, with three determining its overall $SU(2)$ orientation and
one specifying its radius.  

An obvious step toward gaining further understanding of these massless
monopoles and their associated non-Abelian clouds would be to investigate
solutions containing larger numbers of monopoles. Solutions corresponding to
one massless and two identical massive monopoles have been studied in $SU(3)$
broken to $SU(2)\times U(1)$ \cite{dancer,irwin} and in  $Sp(4)=SO(5)$, 
also broken to $SU(2)\times U(1)$ \cite{sp4}.  (In the latter case the 
unbroken $SU(2)\times U(1)$ is a different subgroup than that considered in
Ref.~\cite{so5}.)  In both cases the moduli space
metric was found explicitly, but analytic expressions for  the gauge and Higgs
fields could only be found for special configurations.  

The complexity of these solutions is perhaps not surprising if one recalls
the rather nontrivial form of the Atiyah-Hitchin metric for the moduli space of
two identical $SU(2)$ monopoles \cite{atiyah}.  By contrast, the moduli space 
metric for two \cite{taub}, or even an arbitrary number \cite{many},
of distinct fundamental monopoles is 
relatively simple, and
so one might expect the corresponding solutions for the fields to be more
tractable.  As we will see, this is indeed the case.  We consider here
$SU(N)$ solutions comprising $N-1$ distinct fundamental monopoles.   For both
the case of maximal symmetry breaking, where all of the monopoles are 
massive, and the case where the unbroken group is $U(1)\times SU(N-2)\times
U(1)$, where all but two of the monopoles become massless, the gauge and Higgs
fields can be expressed in terms of elementary functions for all values of
the collective coordinates.

We use Nahm's method to construct these solutions \cite{nahm}.  
In Sec.~II we review the
details of this construction for BPS monopoles in an $SU(N)$ theory.  The
implementation of the construction for configurations containing many distinct
fundamental monopoles is described in Sec.~III.  In Sec.~IV we consider the
case where the unbroken group is $U(1)\times SU(N-2)\times U(1)$ and obtain
explicit expressions for the fields.  
These expressions simplify considerably in the regions outside the cores of
the massive monopoles.   We discuss these asymptotic forms in Sec.~V. 
Section~VI contains some concluding remarks.  There is an Appendix containing
details of some of the calculations.

\section{The Nahm construction}

The fundamental elements in Nahm's construction \cite{nahm}
of the BPS monopole solutions
are a triplet of matrices $T_a(t)$, the Nahm data, that satisfy a set of
nonlinear ordinary differential equations.  These $T_a$ then define a linear
differential equation for a second set of matrices, $v(t, {\bf r})$, from which
the fields $\Phi({\bf r})$ and ${\bf A}({\bf r})$ can be constructed.    
In this section we review the details of this construction for the case
of an $SU(N)$ theory \cite{group} with the asymptotic Higgs field and
magnetic charge given 
by Eqs.~(\ref{phievalues}) and (\ref{Qevalues}),
respectively\footnote{We will in general follow the notation of
Ref.~\cite{bowman}.}. 
Here, and for the remainder of the paper, we set the gauge
coupling equal to unity.

    The matrices $T_a(t)$ are defined for $t_1<t<t_N$.  The $t_j$ divide
this range into $N-1$ intervals.  On the $j$th interval, $t_j<t<t_{j+1}$, we
define $k(t) = n_j$ and require that the $T_a$ have dimension $k(t) \times
k(t)$.  In addition, whenever two adjacent intervals have the same value
for $k(t)$, there are three matrices $\balpha_j$, of dimension $k(t_j) \times
k(t_j)$, defined at the interval boundary $t_j$.  These matrices satisfy
the Nahm equation
\begin{equation} 
   {dT_a \over dt} = {i\over 2}\epsilon_{abc}[T_b,T_c] + \sum_j (\alpha_j)_a
             \delta(t-t_j) \, .
\label{Nahmeq}
\end{equation}
where the sum in the last term (and similar sums in later equations)
should be understood to run only over those values of $j$ such that
$n_j=n_{j-1}$.  (The $T_a$ are singular at $t_{j+1}$ if
$|n_{j+1}-n_j| \ge 2$.  Because we will not be considering such
situations here, we will not describe the requirements obeyed by these
singularities.)

   Having found the Nahm data, the next step is to find a $2k(t) \times N$
matrix function $v(t,{\bf r})$ and $N$-component row vectors $S_j({\bf r})$
obeying the differential equation 
\begin{equation}
 0= \left[ -{d\over dt} + ({ T_a} + {r_a})\otimes { \sigma_a} \right]v 
  + \sum_j a_j^\dagger S_j \delta(t-t_j)
\label{veq}
\end{equation}
and the normalization condition 
\begin{equation} 
  I = \int dt\,  v^\dagger  v    + \sum_j S^\dagger_j S_j  \, .
\label{normalization}
\end{equation}
Here $a_j$ is a $2k(t_j)$-component row vector obeying 
\begin{equation} 
     a^\dagger_j  a_j   = {\balpha}_j\cdot { \bsigma} -i (\alpha_j)_0 I
\label{adef}
\end{equation}
with $(\alpha_j)_0$  chosen so that the above matrix has rank 1.

Finally, the spacetime fields are given by 
\begin{equation}
  \Phi =  \int dt \,t \,v^\dagger  v  + \sum_j t_j
            S^\dagger_j S_j 
\label{phieq}
\end{equation} 
\begin{eqnarray} 
  {\bf A} &=& -i\int dt v^\dagger {\bf \nabla} v    
       -i \sum_j S^\dagger_j {\bf \nabla} S_j  \cr
      &=& -{i\over 2} \int dt \left[v^\dagger  {\bf \nabla} v 
           - {\bf \nabla} v^\dagger \,  v  \right]
         -{i\over 2}  \sum_j \left[ S^\dagger_j {\bf \nabla} S_j 
           -{\bf \nabla} S^\dagger_j S_j \right] 
\label{Aieq}
\end{eqnarray}
where the second equality in Eq.~(\ref{Aieq}) is obtained with the aid of the
normalization condition Eq.~(\ref{normalization}).  These satisfy the self-dual
BPS equations
\begin{equation}
     {\bf B} = {\bf D}\Phi
\end{equation}
where 
\begin{equation}
     B_a ={1\over 2}\epsilon_{abc}F_{bc}=  {1\over 2}\epsilon_{abc}
    \left(\partial_b A_c -\partial_c A_b +i[A_b,A_c] \right)
\end{equation}
\begin{equation}
    D_a\Phi = \partial_a \Phi + i[A_a,\Phi]  \, .
\end{equation}

Equations (\ref{veq}) and (\ref{normalization}) do not completely determine
$v$ and the $S_j$.  Given any solution of these equations, 
a second solution can
be obtained by multiplication on the right by an $N \times N$ unitary matrix
function of $\bf r$; this corresponds to an ordinary gauge transformation.   In
addition, there is also some freedom to multiply $v$ and the $S_j$ on the left,
with corresponding transformations on the Nahm data.  Such transformations have
no effect on the spacetime fields, but can be used to simplify the intermediate
calculations, as we will see in Sec.~4.

\section{Construction of $(1,1,\dots,1)$ monopole solutions}

We will be concerned in this paper with solutions consisting of $N-1$
distinct fundamental monopoles.  The $n_j$ are then all equal to
unity and $k(t)=1$ for the entire range of $t$.
Since $k(t)$ is unchanged at each of the intermediate $t_j$,
there is an $\balpha_j$ and an $S_j$ for each value of
$j$ from 2 through $N-1$.  
The commutator term vanishes, and so Eq.~(\ref{veq}) is easily solved
to give the piecewise constant solution
\begin{equation}
   {\bf T}(t) = - {\bf x}_j     \, , \qquad  t_j < t < t_{j+1}  \, .
\end{equation}
The ${\bf x}_a$ have a natural interpretation as the positions of the
individual monopoles.  The $a_j$ of Eq.~(\ref{adef}) are
simply two-component row vectors, which we take to be
\begin{equation} 
     a_j = \sqrt{2|{\bf x}_j-{\bf x}_{j-1}|} \left( \cos(\theta/2)
                   e^{-i\phi/2},  
                 \sin(\theta/2) e^{i\phi/2}\right)
\end{equation}
where $\theta$ and $\phi$ specify the direction of the vector
$\balpha_j={\bf x}_{j-1} -{\bf x}_j$.

The next step is to find a $2\times N$ matrix $v(t)$ and a set of $N$-component
row vectors $S_k$ ($k=2,3, \dots, N-1$) that satisfy Eq.~(\ref{veq}).  To this
end, we first define for each interval $t_k \le t \le t_{k+1}$
a function $f_k(t)$, with  
\begin{eqnarray}
     f_1(t) &=& e^{(t-t_2)({\bf r} - {\bf x}_1)\cdot { \bsigma}}  \cr   
     f_k(t) &=& e^{(t-t_k)({\bf r} - {\bf x}_k)\cdot { \bsigma}}    
         f_{k-1}(t_k)   \, , \qquad k >1  \, .
\label{fdef}
\end{eqnarray}
These have been defined so that their values at the endpoints of the
intervals satisfy   
\begin{equation}
    f_k(t_k) = f_{k-1}(t_k)  \equiv g_k, \qquad k=2,3, \dots, N-1 \, .
\end{equation}
An arbitrary solution of Eq.~(\ref{veq}) can then be written in the form
\begin{equation}
     v(t) = f_k(t) \eta_k \, , \qquad  t_k < t < t_{k+1} \, ,
\label{etadef}
\end{equation}
with discontinuities at the intermediate $t_k$ obeying
\begin{equation}
      \eta_k=\eta_{k-1} + g^{-1}_k a_k^\dagger S_k  \, .
\label{discontinuity}
\end{equation}
The normalization condition, Eq.~(\ref{normalization}), becomes
\begin{equation} 
    I = \sum_{j = 2}^{N-1} 
            S_j ^\dagger S_j 
     + \sum_{k=1}^{N-1}  {\eta_k}^\dagger N_k \eta_k  
\label{orthog}
\end{equation}
where
\begin{equation}
     N_k = \int_{t_k}^{t_{k+1}}dt \,f_k^\dagger(t) f_k(t)  \, .
\label{Ndef}
\end{equation}

We will find it convenient to distinguish between the first two and the last
$(N-2)$ columns of $v$ and the $S_k$, labeling the former by Roman 
superscripts
from the beginning of the alphabet and the latter by Greek superscripts that
run from 3 to $N$.   We choose the $v^a$ to be continuous, so that 
\begin{equation}
    S_k^a =0   \, ,   \qquad a=1,2  \, .
\end{equation}
A properly normalized solution for the $v^a$ is then obtained by taking
\begin{equation}
    \eta_k^a = N^{-1/2} \theta^a  \, ,   \qquad a=1,2 \, ,
\end{equation}
where 
\begin{equation}
   N =   \sum_{k=1}^{N-1}  N_k  
\end{equation}
and the $\theta^a$ are the
two-component objects $\theta^1=(1,0)^t$ and $\theta^2=(0,1)^t$.

    Orthogonality of each of the last $N-2$ columns of $v$ with the first two
implies that 
\begin{equation}
     0 = \sum_{k=1}^{N-1} N_k \eta^\mu_k \, ,\qquad \mu=3,4, \dots N \, .
\end{equation} 
Together with the discontinuity Eq.~(\ref{discontinuity}), this gives
\begin{equation}
   \eta_j^\mu= N^{-1}  \sum_{k=1}^{N-1} \sum_{l=2}^{N-1} c_{jkl} N_k 
           g_l^{-1} a_l^\dagger S_l^\mu
\label{longetaeq}
\end{equation}
where
\begin{equation}
       c_{jkl}= \cases{ \ \ \, 1, \quad j\ge l > k \, ,\cr
                           -1,  \quad j < l \le k \, ,\cr
                       \ \ \, 0, \quad {\rm otherwise} \, .}   
\end{equation} 
Substituting these solutions for the $\eta_j^\mu$ into the orthogonality
condition, Eq.~(\ref{orthog}), gives
\begin{equation}
    \delta^{\mu\nu} = \sum_{i,j= 2}^{N-1} 
            {S_i^\mu}^\dagger [ \delta_{ij} + a_i M_{ij}a_j^\dagger ]
             S^\nu_j 
    \equiv \sum_{i,j= 2}^{N-1}  {S_i^\mu}^\dagger K_{ij} S_j^\nu
\label{Scondition}
\end{equation}
where
\begin{equation}
    M_{ij} = \sum_{k,l,m=1}^{N-1} c_{kli}c_{kmj} {g_i^{-1}}^\dagger N_l^\dagger
   N^{-1}    N_k N^{-1} N_m g_j^{-1}\, .
\end{equation}
Hence, 
\begin{equation}
      S_j^\nu = \left( K^{-1/2}\right)_{j\mu} U_{\mu\nu}
\label{Ssolution}
\end{equation}
where $U$ is any $(N-2)\times(N-2)$ unitary matrix\footnote{In applying this
equation, one must be careful to take into account our convention that the
upper index on $S$ runs from 3 to $N$, while the lower runs from 2 to $N-1$.}. 
The freedom to choose $U$ corresponds to a $U(N-2)$ subgroup of the $SU(N)$
gauge symmetry; the remaining gauge symmetry has already been fixed by our
choices for the first two columns of $v$.

     Substituting this expression for the $S_j^\nu$ into Eq.~(\ref{longetaeq})
gives the $\eta_j^\mu$ and thus, through Eq.~(\ref{etadef}), determines
$v(t)$.  It is then a straightforward, although tedious, matter to substitute
these results into Eqs.~(\ref{phieq}) and (\ref{Aieq}) 
and thus obtain the fields
${\bf A}({\bf r})$ and $\Phi({\bf r})$.  We will not carry this out explicitly
for the case of maximal symmetry breaking.  However, we note that it is clear
from the above equations that the result can 
be expressed in terms of elementary
functions.

\section{Solutions for $SU(N) \rightarrow U(1)\times SU(N-2)\times U(1)$}

     Our main interest is in the case where the middle $N-2$ eigenvalues of
the asymptotic Higgs field are all equal, so that the unbroken gauge group is
$U(1) \times SU(N-2) \times U(1)$.  If we adjust the Higgs field in this
fashion, then, as was argued in Ref.~\cite{nonabelian}, the $(N-1)$-monopole 
solutions of the previous section can be viewed as being composed of
two massive and $N-3$ massless monopoles, with the latter condensing into a
non-Abelian ``cloud''.  The massive monopoles are located at ${\bf x}_1$
and ${\bf x}_{N-1}$; without any loss of generality we may take these to  lie
on the $z$-axis, with $z_{N-1} = z_1 +R \ge z_1$.  The
locations of the massless monopoles are less well-defined.  Extrapolating from
the maximally broken case, one would take these to be the
points ${\bf x}_2, {\bf x}_3, \dots, {\bf x}_{N-2}$.  However, as we will
now show, many different choices for these points yield the same
solution.

     To begin, note that of the $N-1$ intervals into which the range of $t$
was divided, only the leftmost ($t_1 \le t \le t_2$) and rightmost 
($t_2 =t_{N-1} \le t \le t_N$) now have nonzero width; we shall use subscripts
$L$ and $R$ to label quantities related to these two intervals.  Hence, of the
$N-1$ integrals defined by Eq.~(\ref{Ndef}), only $N_1 \equiv N_L$ and
$N_{N-1}\equiv N_R$ are nonzero.   Also, the $g_k=f_k(t_k)$ are all equal to
unity.   As a result, Eq.~(\ref{Scondition}) simplifies to 
\begin{equation}
    \delta^{\mu\nu} = \sum_{i,j} {S_i^\mu}^\dagger 
       \left[ \delta_{ij} + a_i M a^\dagger_j \right]
            S^\nu_j  
\label{newScondition}
\end{equation}
where 
\begin{equation} 
     M =  \left(N_L^{-1} +N_R^{-1} \right)^{-1}
\label{Mdef}
\end{equation} 
is Hermitian.
Once a set of $S^\nu_j$ satisfying Eq.~(\ref{newScondition}) has been found,
the $\eta_k$, and hence $v(t)$, can be found from  Eq.~(\ref{longetaeq}),
which now reduces to  
\begin{eqnarray}
   \eta_1^\mu &=& -N_L^{-1} M \sum_j a_j^\dagger S_j^\mu \cr 
   \eta_{N-1}^\mu &=& N_R^{-1} M  \sum_j a_j^\dagger S_j^\mu \, .
\label{etaone}
\end{eqnarray}
  
    With maximally broken symmetry, the monopole positions enter
both through the functions $f_k(t)$ and
through the various $a_j$.  In the present case, where the middle
intervals have zero width, the ${\bf x}_k$ associated with the massless
monopoles enter only through the $a_j$, which, as can be seen from
Eqs.~(\ref{newScondition}) and (\ref{etaone}), appear only in the combination
$\sum_j a_j^\dagger S_j^\mu $.  With this in mind, consider two
sets of monopole positions ${\bf x}_k$ and $\tilde{\bf x}_k$ with
identical locations for the massive monopoles, but with the massless
monopoles constrained only by the requirement that $\tilde a_j =W_{jk}
a_k$, with $W$ some $(N-2)\times (N-2)$ unitary matrix.  If
$S_j^\mu$ gives a solution of Eq.~(\ref{newScondition}) for the former
set of positions, $\tilde S_j^\mu = W_{jk} S_k^\mu$ is a
solution for the transformed set.  (Note that this
transformation has no effect on either $v$ or on the fields
themselves.)  

     The possibility of performing such transformations implies that the 
positions of the massless monopoles are not all physically meaningful
quantities.  In fact, these yield only a single physical parameter, which can
be identified by noting that these transformations leave invariant the quantity
\begin{equation}
    \sum_j  a^\dagger_j a_j = 
 \sum_j
 \left[ {\bf\alpha}_j\cdot { \bsigma} -i {\alpha_j}_0 I \right]
     = ({\bf x}_1 - {\bf x}_{N-1})\cdot \bsigma 
         + \sum_{j=2}^{N-1} |{\bf x}_j - {\bf x}_{j-1}| \, .
\end{equation} 
The first term on the right hand side is fixed by the positions of the two
massive monopoles.  The second term is just the sum of the distances between
successive monopoles.  It is precisely this sum that was identified in 
Ref.~\cite{nonabelian}, from the properties of the moduli space, as the unique
gauge-invariant quantity characterizing the non-Abelian cloud.  It will be
convenient to express it in terms of the ``cloud parameter'' $b\ge 0$, defined
by the equation
\begin{equation}
     2b +R = \sum_{j=2}^{N-1} |{\bf x}_j - {\bf x}_{j-1}| \, .
\label{distancesum}
\end{equation}

     It is not hard to show that any two sets of massless monopole positions
corresponding to the same value of $b$ can be transformed into one another.  
We may therefore define a canonical set of positions by placing one of the
massless monopoles on the $z$-axis at $z_2=z_1-b$ and the remaining $N-3$
massless monopoles on top of the massive monopole at $z_{N-1}$.  The only
nonvanishing $a_j$ are $a_2$ and $a_3$, which can be combined as the
rows of a real $2 \times 2$ matrix
\begin{equation}
    {\cal A} = \left(\matrix{ \sqrt{2b} & 0 \cr 0 & \sqrt{2(b+R)}} \right) \, .
\label{amatrix}
\end{equation}   

    With this canonical choice of monopole positions, the 
solution for the $S_k$ given in Eq.~(\ref{Ssolution}) takes on a particularly
simple form.  If   these $N-2$ row vectors are combined to form an $(N-2)\times
N$ matrix $S$, the solution corresponding to the choice $U=I$ can be written as
\begin{equation}
    S  =  \left(\matrix{0 &  {\cal S} &  0 \cr
                   0&  0&  I_{N-4}  }\right) 
\end{equation}
where the columns (rows) have been grouped in blocks of 2, 2, and
$N-4$ (2 and $N-4$) and the 
$2\times 2$ matrix 
\begin{equation}
     {\cal S} =  {\cal S}^\dagger = (I +  {\cal A}M {\cal A} )^{-1/2} \, .
\end{equation}

     Equations~(\ref{etadef}) and (\ref{etaone}) then give $v(t)$, which in a
similar block notation takes the form
\begin{equation}
    v(t) = \cases{ f_1(t) \left( N^{-1/2}, - N_L^{-1}(ML)^{1/2}, 0 \right)
                     \, , \quad t_1 \le t \le t_2  \, ,\cr 
                 f_{N-1}(t) \left( N^{-1/2},  N_R^{-1}(ML)^{1/2}, 0 \right)
                \, , \quad t_2 \le t \le t_N  \, .}
\end{equation}
Here $N = N_L+N_R$ and we have defined
\begin{equation}
      L = M  {\cal A} ^2 (I +  {\cal A}M {\cal A} )^{-1} \, .
\label{Ldef}
\end{equation}
Note that, as we will see explicitly below, $\cal A$, $M$, and $L$ are all diagonal
and hence commute.
     
    These results can now be combined to yield the gauge and Higgs
fields.  To express these, it is convenient to introduce the integrals 
\begin{eqnarray}
   {\bf H}_L &=& \int_{t_1}^{t_2}dt  [f_1(t),{\bf \nabla} f_1(t)] \cr 
    K_L &=& \int_{t_1}^{t_2}dt \,t \,f_1(t)^2  -t_2 N_L 
\end{eqnarray}
and the similarly defined quantities $K_R$ and ${\bf H}_R$ involving
$f_{N-1}(t)$.   The fields can then be written as
\begin{equation}
     {\bf A} =  \left(\matrix{N^{-1/2} & 0 & 0 \cr
                   0 & (LM)^{1/2}  & 0 \cr
                    0 & 0 & 0  }\right) 
              \left(\matrix{{\bf a}^{(1)} &{\bf a} ^{(3)} & 0 \cr
                   {{\bf a}^{(3)}}^\dagger &{\bf a} ^{(2)}  & 0 \cr
                    0 & 0 & 0  }\right) 
               \left(\matrix{N^{-1/2} & 0 & 0 \cr
                   0 & (ML)^{1/2}  & 0 \cr
                    0 & 0 & 0  }\right) 
\label{bigai}
\end{equation}      
\begin{equation}
     \Phi =  t_2I  + \left(\matrix{N^{-1/2} & 0 & 0 \cr
                   0 & (LM)^{1/2}  & 0 \cr
                    0 & 0 & 0  }\right) 
              \left(\matrix{\phi^{(1)} & \phi^{(3)} & 0 \cr
                   {\phi^{(3)}}^\dagger & \phi^{(2)}  & 0 \cr
                    0 & 0 & 0  }\right) 
               \left(\matrix{N^{-1/2} & 0 & 0 \cr
                   0 & (ML)^{1/2}  & 0 \cr
                    0 & 0 & 0  }\right) 
\label{bigphi}
\end{equation}  
where the ${\bf a}^{(a)}$ and $\phi^{(a)}$ are the $2\times 2$ matrices
\begin{eqnarray}
     {\bf a}^{(1)} &=& -{i\over 2}   \left( {\bf H}_L +{\bf H}_R
                  \right)   +{i\over 2} \left[N^{1/2}, {\bf
                  \nabla} N^{1/2}\right] \cr 
    \phi^{(1)} &=&   (K_L +K_R)   \cr 
    {\bf a}^{(2)} &=& -{i\over 2}  ( N_L^{-1} {\bf H}_L
           N_L^{-1} + N_R^{-1}
           {\bf H} _R N_R^{-1} )  \cr 
       \phi^{(2)} &=&  N_L^{-1} K_L N_L^{-1} + N_R^{-1} K_R
                  N_R^{-1}  \cr 
      {\bf a}^{(3)} &=& {i\over 2}[ ({\bf \nabla} N_R -{\bf
            H}_R) N_R^{-1} -  ({\bf \nabla} N_L -{\bf H}_L) N_L^{-1}]
                 \cr 
        \phi^{(3)} &=&   K_R N_R^{-1} -K_L  N_L^{-1} \, .
\label{generalfields}
\end{eqnarray}

To proceed further we need explicit expressions for the various
integrals that we have defined. These are most easily expressed in terms
of the vectors
\begin{eqnarray}
   {\bf y}_L &=& {\bf r} - {\bf x}_1  \cr
   {\bf y}_R &=& {\bf r} - {\bf x}_{N-1}
\end{eqnarray}
and the quantities
\begin{eqnarray}
    s_L &=& (t_2-t_1)|{\bf y}_L| = (t_2-t_1)y_L  \cr
    s_R &=& (t_N-t_2)|{\bf y}_R| = (t_N-t_2)y_R  \, .
\end{eqnarray}
Straightforward integration yields
\begin{eqnarray}
    N_L &=& y_L^{-1} \sinh s_L e^{-s_L {\hat {\bf y}}_L \cdot  \bsigma} \cr
    N_R &=& y_R^{-1} \sinh s_R e^{s_R  {\hat {\bf y}}_R \cdot
           \bsigma} \cr 
    K_L &=& {1\over 2} y_L^{-1} {\hat {\bf y}}_L\cdot \bsigma 
         \left[(t_2-t_1)  e^{-2s_L {\hat {\bf y}}_L \cdot  \bsigma} 
          -N_L\right] \cr
    K_R&=& {1\over 2} y_R^{-1} {\hat {\bf y}}_R\cdot \bsigma 
         \left[(t_N-t_2)  e^{2s_R {\hat {\bf y}}_R \cdot  \bsigma} 
          -N_R\right] \cr
    {\bf H}_L &=& -i  y_L^{-2} ({\hat {\bf y}}_L \times
          \bsigma)  
    \left( \sinh s_L \cosh s_L - s_L \right) \cr 
     {\bf H}_R &=& -i  y_R^{-2} ({\hat {\bf y}}_R \times
          \bsigma) 
    \left( \sinh s_R \cosh s_R - s_R \right)  \, .
\end{eqnarray}
    
Substituting these expressions into Eq.~(\ref{generalfields}) yields the
remarkably simple formulas 
\begin{eqnarray}
    {\bf a}^{(2)} &=&-{1\over 2}  ({\bf V}\times \bsigma)    \cr
    \phi^{(2)} &=& -{1\over 2}  ({\bf V}\cdot\bsigma)      \cr
    {\bf a}^{(3)} &=& {i\over 2}  F \bsigma    \cr
    \phi^{(3)} &=& {1\over 2}  F     
\label{VandFform}
\end{eqnarray}
where 
\begin{equation}
   {\bf V} = {\hat {\bf y}}_L \left[ \coth s_L - {s_L\over \sinh^2s_L}  \right]
      +{\hat {\bf y}}_R \left[ \coth s_R - {s_R\over \sinh^2s_R}  \right]
\end{equation}
and 
\begin{equation}
    F= {{\hat {\bf y}}_L\cdot \bsigma\over y_L}
      - {{\hat {\bf y}}_R\cdot \bsigma\over y_R}
      +(t_2-t_1)\left(1-{\hat {\bf y}}_L\cdot \bsigma 
       \coth s_L \right)
      +(t_N-t_2)\left(1+{\hat {\bf y}}_R\cdot \bsigma
       \coth s_R \right)  \, .
\end{equation}

Combining the expressions for $N_L$ and $N_R$ with Eq.~(\ref{Mdef}), and
using the fact that 
\begin{equation}
 ({\bf y}_L - {\bf y}_R)\cdot \bsigma  = R \sigma_3
\end{equation}
we obtain
\begin{equation}
  M  = \left( y_L \coth s_L + y_R \coth s_R + R \sigma_3 \right)^{-1} \, .
\end{equation}
Combining this with
Eqs.~(\ref{amatrix}) and (\ref{Ldef}) yields
\begin{equation}
  L  =  { 2b +R - R\sigma_3   \over  
       y_L \coth s_L + y_R \coth s_R + R +2b}   \, .
\label{Lexpression}
\end{equation}
Note that the cloud parameter $b$ enters the solution only through the
matrix $L$. 

\section{Asymptotic behavior} 

\def\bhyl{\hat{\bf y}_L}
\def\bhyr{\hat{\bf y}_R}
\def\hyl{\hat{ y}_L}
\def\hyr{\hat{ y}_R}

In this section we will examine in some detail the solutions that we
have found.  From the form of Eqs.~(\ref{bigai}) and (\ref{bigphi}),
it is clear that for any $N\ge 5$ all $SU(N)$ solutions are essentially
embeddings\footnote{For the special case $R=0$ and $t_1=-t_N$, these
actually reduce to embeddings of the $SO(5)$ solution of Ref.~\cite{so5}}
 of solutions with one massless and two massive
monopoles in $SU(4)$ broken to $U(1)\times SU(2)\times U(1)$.  Therefore,
without any loss of generality we can
simplify our notation by specializing to the case $N=4$.  Each
adjoint representation elementary multiplet of the theory can then be
decomposed into five massless fields (an $SU(2)$ triplet and two
singlets) together with a pair of massive doublets and a pair of
massive singlets.  Because of the ``twisting'' of the topologically
nontrivial Higgs field, this decomposition will not in general have a
simple correspondence with the matrix components of the fields.
However, one might hope that matters would simplify in the region
outside the cores of the massive monopoles (i.e., the region where
$s_L$ and $s_R$ are both much greater than unity), where the massive
fields would be expected to be exponentially small.

The first hint of this simplification comes from noting that $N_L$ and
$N_R$ each have one exponentially large eigenvalue.  Hence, the
eigenvalues of $N=N_L+N_R$ are in general both exponentially large, implying
that those of $N^{-1/2}$ are both exponentially small.  The only exception
to this occurs when the large-eigenvalue eigenvectors of $N_L$ and $N_R$ are
almost parallel to each other, which happens only near the line joining the
centers of the two massive monopoles.  If we exclude the region close to this
intermonopole axis, no elements of the matrices $F$, $M$, and $L$ are ever
exponentially large.  We then immediately see from
Eq.~(\ref{generalfields}) that the terms containing ${\bf a}^{(3)}$
and $\phi^{(3)}$ are exponentially small.

If we therefore restrict ourselves to the region where $s_L, s_R \gg
1$ (so that we are outside the massive cores) and 
\begin{equation}
   {y_L +y_R -R \over R}  \gg  e^{-2s_L} + e^{-2s_R}
\label{axisregion}
\end{equation}
(to avoid the intermonopole axis), we may approximate the fields by
the block diagonal form
\begin{equation}
     {\bf A} = \left(\matrix{{\bf A}^{(1)} & 0   \cr
                   0 & {\bf A}^{(2)}  }\right) 
\label{Ablock}
\end{equation}      
\begin{equation}
     \Phi =    \left(\matrix{\Phi^{(1)} & 0   \cr
                   0 & \Phi^{(2)} }\right) \, .
\label{Phiblock}
\end{equation}
With the fields written in this form, their group theoretic
interpretation is fairly clear.  The traceless parts of the
nonvanishing blocks correspond to two commuting $SU(2)$ subgroups, one
of which (the lower right, as we shall see) must be the unbroken
$SU(2)$.  The massless $U(1)$ fields are then contained in the
traceless part of the other block and in the two traces.  Examining
Eqs.~(\ref{bigai}) and (\ref{bigphi}) and recalling that the
dependence on the cloud parameter is only through $L$, we see that
${\bf A}^{(2)}$ and $\Phi^{(2)}$ may be $b$-dependent, but ${\bf
A}^{(1)}$ and $\Phi^{(1)}$ are not.  (Because the fields are traceless, the
$b$-dependence must be entirely in the traceless parts of ${\bf A}^{(2)}$ 
and $\Phi^{(2)}$.)

Because of the factors of $N^{-1/2}$, the analysis required to obtain
the asymptotic form for the upper left block is in general somewhat
tedious.  However, the calculation simplifies considerably if
$e^{2s_L}/y_L$ is either much less than or much greater than
$e^{2s_R}/y_R$ (which is the case in almost of space.)  If we
define the unit vector $\hat{\bf n}$ to be equal to
$\bhyr$ ($\bhyl$) in the former (latter) region, then, up to
exponentially small corrections, the gauge and Higgs fields are
\begin{equation}
     {\bf A}^{(1)} = {(y_L+y_R) \over 2[(y_L+y_R)^2 -R^2]}
      \left\{  (\bhyl\times\bhyr) + [\hat{\bf n} \times 
     (\bhyl + \bhyr)] \hat{\bf n}\cdot \bsigma \right\}
        + {i\over 4} \left[ {\bf \nabla} \hat{\bf n}\cdot \bsigma, 
                     \hat{\bf n}\cdot \bsigma \right]
\label{asymA1}
\end{equation}
and
\begin{equation}
      \Phi^{(1)} =  \left(t_4 -{1\over 2y_R}\right)
	   \left({ 1+\hat{\bf n}\cdot \bsigma \over 2}\right)
           + \left(t_1 +{1\over 2y_L}\right) 
             \left({ 1-\hat{\bf n}\cdot \bsigma \over 2}\right)
\end{equation}
while the asymptotic field strength is 
\begin{equation}
    {\bf B}^{(1)} =  {\bhyr\over 2y_R^2}  
	   \left({ 1+\hat{\bf n}\cdot \bsigma \over 2}\right)
             - {\bhyl\over 2y_L^2} 
         \left({ 1-\hat{\bf n}\cdot \bsigma \over 2}\right) \, .
\label{asymB1}
\end{equation}
(In fact, it is not hard to show that $\Phi^{(1)}$ and ${\bf B}^{(1)}$
must be of this form (although with a more complicated expression for
$\hat {\bf n}$) at all points outside the massive cores.)

Although the calculation of the fields in the lower right corner is in
principle straightforward, somewhat lengthy manipulations needed to
put the result in a simple form.  We leave the details of these to the
Appendix, and state the results here.  From
Eqs.~(\ref{bigai}) and (\ref{VandFform}) we obtain  
\begin{equation}
    {\bf A}^{(2)} = -{1\over 2}(LM)^{1/2} {\bf V}\times \bsigma
       (ML)^{1/2}   
\end{equation}
where we can now use the asymptotic forms
\begin{equation}
    {\bf V} = \bhyl +\bhyr    
\label{asymV}
\end{equation}
\begin{equation}
     M = (y_L +y_R +R\sigma_3)^{-1}
\end{equation}
and 
\begin{equation}
  L  = { 2b +R - R\sigma_3   \over  
     y_L + y_R + R +2b} \equiv h ( 2b +R - R\sigma_3)  \, .
\label{asymL}
\end{equation}
For the Higgs field we obtain the particularly simple form
\begin{equation}
    \Phi^{(2)} = t_2 I + L^{1/2}\left[ 
   {1\over 2y_R} \left({1- \hat{\bf q}\cdot\bsigma  \over 2}\right)
  - {1\over 2y_L} \left({1+ \hat{\bf q}\cdot\bsigma  \over 2}\right)
        \right] L^{1/2}
\label{phi2asym}
\end{equation}
where $\hat{\bf q}$ is a unit vector with components
\begin{equation}
   {\hat q}_a  = \cases{ {(y_L)_a + (y_R)_a \over \sqrt{(y_L+y_R)^2 -R^2}}
            \, , \qquad a=1,2 \, ,\cr  \cr
         {y_L - y_R \over R}   \, , \qquad a=3 \, .} 
\label{qcomponents}
\end{equation}
(Note that at large distances $\hat{\bf q}$, like $\hat{\bf n}$, approaches the
radial unit vector.)  The field strength is 
\begin{equation}
   {\bf B}^{(2)}  = L^{1/2}\left\{ \left[{\bhyl\over 2y_L^2} 
           + h{(\bhyl+\bhyr)\over y_L} \right]
           \left({1+ \hat{\bf q}\cdot\bsigma  \over 2}\right)
     - \left[{\bhyr\over 2y_R^2} 
           + h{(\bhyl+\bhyr)\over y_R} \right]
           \left({1- \hat{\bf q}\cdot\bsigma  \over 2}\right)
     - {h\over 2 y_Ly_R} {\bf f} \right\}  L^{1/2}
\label{b2asym}
\end{equation}
where $h$ is defined by Eq.~(\ref{asymL}) and 
\begin{equation}
   f_a = \cases{ \sqrt{(y_L+y_R)^2 -R^2} \, \sigma_a \, , \qquad a=1,2 \, ,\cr
\cr 
              (y_L +y_R) \, \sigma_3 -R \, , \qquad a=3  \, .}
\label{fcomponents}
\end{equation}

It is instructive to consider several limiting cases.  First, suppose
that $b=0$. From Eq.~(\ref{Lexpression}) we see that $L$ is then
proportional to $1-\sigma_3$ so that, except for a constant term in
$\Phi$, the third rows and columns of all fields vanish.  The
solutions are then essentially embeddings of $SU(3)\rightarrow
U(1)\times U(1)$ solutions \cite{athorne}.  Since the unbroken
subgroup of $SU(3)$ is Abelian, it should be possible to choose a
gauge in which the asymptotic fields are simply superpositions of
single monopole fields.  Indeed, these can be written in the form
\begin{equation}
    \Phi = U_1^{-1}({\bf r})
        \left( \matrix{ t_4 -{1\over 2y_R} & 0 & 0 & 0 \cr\cr
             0 & t_2 -{1\over 2y_L} + {1\over 2y_R} & 0 & 0 \cr\cr
             0 & 0 & t_2 & 0 \cr\cr
             0 & 0 & 0 &  t_1 +{1\over 2y_L}} \right) 
        U_1({\bf r})
\end{equation}
\begin{equation}
    {\bf B} = U_1^{-1}({\bf r})
        \left( \matrix{  {\bhyr\over 2y_R^2} & 0 & 0 & 0 \cr\cr
             0 & {\bhyl\over 2y_L^2} - {\bhyr\over 2y_R^2} & 0 & 0 \cr\cr
             0 & 0 & 0 & 0 \cr\cr
             0 & 0 & 0 &   -{\bhyl\over 2y_L^2}    } \right) 
        U_1({\bf r})  \, .
\label{b0asymB}
\end{equation}
[Here, in order to make the $U(1)\times SU(2)\times U(1)$ structure of
the theory clearer, we have reordered the rows and columns to
correspond with the order in Eqs.~(\ref{phievalues}) and
(\ref{Qevalues}).  With this reordering, the unbroken $SU(2)$ is
contained in the middle $2\times 2$ block, and a fundamental monopole
solution with a single nonzero $n_j$ corresponds to an embedding in a
pair of adjacent rows and columns.]  Viewed as an $SU(3)$ solution,
this corresponds to a configuration containing one each of the two
massive fundamental monopoles of the theory, located at points ${\bf
x}_1$ and ${\bf x}_3$.  Viewed as an $SU(4)$ solution, it can be
interpreted as containing a massive fundamental monopole with
$n_j=\delta_{j1}$ at ${\bf x}_1$ and a superposition of a massive
monopole with $n_j=\delta_{j3}$ and a massless
monopole\footnote{Viewing the massless monopole as being centered at
${\bf x}_3$ is a gauge choice; moving it to ${\bf x}_1$ simply
corresponds to a re-ordering of the rows and columns.} with
$n_j=\delta_{j2}$ at ${\bf x}_3$.  Even though the underlying $SU(3)$
solution has purely Abelian long-range fields, the long-range part of
the $SU(4)$ solution is non-Abelian in the sense that the unbroken
$SU(2)$ subgroup acts nontrivially on ${\bf A}^{(2)}$ and
$\Phi^{(2)}$.  The crucial point, however, is that the $SU(2)$
orientations of the Coulomb field centered at ${\bf x}_1$ and the
Coulomb field centered at ${\bf x}_3$ are aligned, so that the
net $SU(2)$ component is a purely dipole field that falls as $R/y^3$
for $y \equiv (y_L + y_R)/2 \gg R$.

Next, consider the case $b \gg R$.  Now $L$ is approximately
proportional to a unit matrix, with 
\begin{equation}
       L = { 2b   \over   y_L + y_R + 2b}   +
        O(R/b) \, ,
\end{equation}
while $h \approx 1/(2b +y_L +y_R)$.  Thus, in the region where $y_L$
and $y_R$ are much less than $b$ the fields can be written as
\begin{equation}
    \Phi = U_2^{-1}({\bf r})
        \left( \matrix{  t_4  - {1\over 2y_R} & 0 & 0 & 0 \cr\cr
             0 & t_2+{1\over 2y_R} & 0 & 0 \cr\cr
             0 & 0 & t_2 -{1\over 2y_L} & 0 \cr\cr
             0 & 0 & 0 & t_1 +{1\over 2y_L} } \right) 
        U_2({\bf r})    + \cdots
\end{equation}
\begin{equation}
    {\bf B} = U_2^{-1}({\bf r})
        \left( \matrix{ {\bhyr\over 2y_R^2} & 0 & 0 & 0 \cr\cr
             0 &  -{\bhyr\over 2y_R^2} & 0 & 0 \cr\cr
             0 & 0 &{\bhyl\over 2y_L^2} & 0 \cr\cr
             0 & 0 & 0 &  -{\bhyl\over 2y_L^2}   } \right) 
        U_2({\bf r}) + \cdots
\end{equation}
where the dots represent terms
that are suppressed by powers of $R/b$, $y_L/b$, or $y_R/b$.  
These are just the fields expected for two massive monopoles, with topological
charges $n_j=\delta_{j1}$ and $n_j=\delta_{j3}$, centered at ${\bf x}_1$ and
${\bf x}_3$, respectively.  In contrast with the previous case, the $SU(2)$
components of their two magnetic charges are not aligned, and so the  unbroken
$SU(2)$ contains two Coulomb fields rather than the dipole field of
Eq.~(\ref{b0asymB}). These non-Abelian Coulomb fields disappear 
when $y \gg b$, where we obtain
\begin{equation}
    \Phi = U_3^{-1}({\bf r})
        \left( \matrix{ t_4 -{1\over 2y} & 0 & 0 & 0 \cr\cr
             0 & t_2 & 0 & 0 \cr\cr
             0 & 0 & t_2  & 0 \cr\cr
             0 & 0 & 0 & t_1 +{1\over 2y}  } \right) 
        U_3({\bf r})    + O(b/y^2)
\end{equation}
\begin{equation}
    {\bf B} = U_3^{-1}({\bf r})
        \left( \matrix{  {\hat{\bf y}\over 2y^2} & 0 & 0 & 0 \cr\cr
             0 & 0 & 0 & 0 \cr\cr
             0 & 0 & 0 & 0 \cr\cr
             0 & 0 & 0 &  -{\hat{\bf y}\over 2y^2}} \right) 
        U_3({\bf r}) +  O(b/y^3) \, .
\end{equation}
Thus, at distances large compared to $b$ the magnetic fields exhibit the
behavior that would be expected if a single massless monopole were added to
the two massive fundamental monopoles.  However, this massless monopole is not
manifested as a localized structure with a well-defined center.  Instead, we
simply have a transition from a ``cloud'' region of size $\sim b$ containing
non-Abelian Coulomb magnetic fields to an outer region where these fields are
cancelled. 

In this discussion we have excluded the region close to the
intermonopole axis.  To explore the fields in this region, one must go
back to the equations of the previous section.
[Equations~(\ref{asymV}--\ref{asymL}) are not valid approximations here, even
outside the monopole cores.] Doing so, we find that the fields do not
have the simple block diagonal structure of Eqs.~(\ref{Ablock}) and
(\ref{Phiblock}).  In
addition, some components of $\bf A$ become exponentially large as one
approaches the axis, while $\Phi$ turns out to be rapidly varying.
However, these are essentially artifacts of our choice of gauge.  The
net magnetic charge of our solutions is a unit charge in the unbroken
$U(1)$ that is contained in ${\bf A}^{(1)}$.  The long-range twisting of
the Higgs field must then be topologically equivalent to an embedding of
the $SU(2)$ hedgehog configuration in the corresponding $SU(2)$; this can be
seen in the behavior of the unit vector $\hat{\bf n}$ that appears in
Eqs.~(\ref{asymA1}--\ref{asymB1}).  However, this cannot be the whole story.
Because the two monopoles have different $U(1)$ charges, there must be some 
additional twisting of the Higgs field near each of the monopoles. The
conventions that we have adopted are such that this inevitable additional
twisting is confined to the narrow region near the axis where
Eq.~(\ref{axisregion}) does not hold.  In order that ${\bf D}\Phi$ and $\bf B$
not become large, these rapid variations in the direction of $\Phi$ must be
compensated by large values of $\bf A$.  That this actually happens can be
verified by evaluating the field strength along the axis.  One finds that, up
to exponentially small corrections, the magnetic field along this axis is
independent of $b$ and can be put in the form of Eq.~(\ref{b0asymB}).

\section{Concluding Remarks} 

In this paper we have shown how the Nahm construction can be used to
obtain explicit multimonopole solutions corresponding to $N-1$
distinct fundamental monopoles in $SU(N)$.  In the case where the
symmetry is broken maximally, to $U(1)^{N-1}$, these solutions are
described by $3(N-1)$ gauge-invariant parameters that specify the
positions of the component monopoles; these combine with $N-1$ overall
$U(1)$ phases to give the full set of collective coordinates.  Even
though these solutions have no rotational symmetry at all, the gauge
and Higgs fields can be expressed in terms of elementary (i.e.,
rational and hyperbolic) functions for arbitrary values of these
parameters.  Of particular interest is the behavior of these solutions
in the limit where the unbroken group is enlarged to $U(1)\times
SU(N-2)\times U(1)$ and $N-3$ of the component monopoles become
massless.  Examining the solutions, one sees no trace of the
individual massless monopoles, but only a single ``non-Abelian
cloud''.  Indeed, most of the massless monopole coordinates become
redundant in this limit, with the solutions depending only on a sum of
intermonopole distances that is conveniently described by the cloud
parameter $b$.  Not only are the massless monopole positions somewhat
ill-defined, but so, in a sense, is their number.  As we have seen,
the $SU(N)$ solution that nominally contains $N-3$ massless monopoles
is essentially equivalent to an embedded $SU(4)$ solution containing a
single massless monopole.  

For values of $b$ that are large compared to the distance $R$ between
the massive monopoles, the cloud is rather similar to that found in
the $SO(5)$ case.  The non-Abelian Coulomb magnetic fields due to the
two massive monopoles are present inside the cloud, but are screened
at distances much larger than $b$.  On the other hand, when $b\ll R$
these Coulomb fields are present outside the cloud (i.e., at points
more than a distance $b$ from either of the massive cores), but are
aligned with each other in such a manner that the entire solution is
simply an embedding of a purely Abelian configuration.

These non-Abelian clouds and their properties clearly warrant further
study.  For the configurations we have considered here, those parts
of the solution associated with the cloud are particularly simple.
The dependence on the cloud parameter is contained in a single matrix
$L$, while corresponding components of the Higgs and gauge fields
depend on the same function of the spatial variables and differ only
in their tensor structure.  In addition, the actual functional forms
outside the massive cores are relatively simple.  This simplicity
suggests that it might be feasible to determine the structure of the
non-Abelian clouds associated with more complex configurations, at
least in the regions outside the cores of the massive monopoles.  The
most compelling questions are associated with the conjectured
electric-magnetic duality.  From this point of view, the massless
monopoles are clearly the duals of the massless non-Abelian gauge
bosons and their superpartners.  How do these particles reflect the
strange properties of the massless monopoles?

We are grateful to Choonkyu Lee, Kimyeong Lee and Changhai Lu for helpful
conversations.  We thank the Aspen Center for Physics, where this work
was begun.  This work was supported in part by the U.S. Department of
Energy and the U.S. National Science Foundation.

\vskip 1cm
\setcounter{equation}{0}
\makeatletter
\renewcommand\theequation{A.\arabic{equation}}
\makeatother
 
\leftline{\Large\bf Appendix}
\vskip 5mm
\noindent
In this Appendix we outline the manipulations that lead to Eqs.~(\ref{phi2asym}) and
(\ref{b2asym}) for $\Phi^{(2)}$ and ${\bf B}^{(2)}$.  Throughout, $\bf V$, $M$,
and $L$ should be understood to be given by the asymptotic forms in
Eqs.(\ref{asymV}-\ref{asymL}).  

The first step is to obtain some useful identities.  With the aid of the law of
cosines, we obtain
\begin{equation}
     {\bf V}^2 = 2(1+ {\hat{\bf y}}_L \cdot {\hat{\bf y}}_R )
      =   {(y_L +y_R)^2 - R^2 \over y_L y_R }  
\end{equation}
Because we have chosen the massive monopoles to lie along the $z$-axis, the third
component of any vector can be obtained by noting that $Rw_3= {\bf w}\cdot
({\bf y}_L - {\bf y}_R)$.  In particular, many of the subsequent results make
use of the identities 
\begin{equation}
     ({\hat y}_L)_3  
       =  { (y_L+y_R)\over R} -{(y_L +y_R)^2 - R^2 \over 2Ry_L }  
\label{yLthree}
\end{equation}
and
\begin{equation}
     ({\hat y}_R)_3  
       =  -{ (y_L+y_R)\over R} +{(y_L +y_R)^2 - R^2 \over 2Ry_R }  
\label{yRthree}
\end{equation}

Next, we define a vector $\bf f$ by the equation
\begin{equation}
     M^{1/2}  \bsigma M^{1/2} = \left[ (y_L +y_R)^2 - R^2  \right]^{-1} {\bf f}
\label{MsigmaM}
\end{equation}
Explicit expressions for the components of $\bf f$ are given in
Eq.~(\ref{fcomponents}).  The inner products of this equation with ${\hat{\bf
y}}_L$ and ${\hat{\bf y}}_R$ then lead, with the aid of Eqs.~(\ref{yLthree}) and
(\ref{yRthree}), to \begin{equation}
   M^{1/2} {\hat{\bf y}}_L \cdot \bsigma M^{1/2} 
      = \left( {{\hat{\bf q}} \cdot \bsigma + 1 \over 2y_L }\right) -M
\end{equation}
and
\begin{equation}
   M^{1/2} {\hat{\bf y}}_R \cdot \bsigma M^{1/2} 
      = \left( {{\hat{\bf q}} \cdot \bsigma - 1 \over 2y_R }\right) +M
\label{MyrM}
\end{equation}
where $\hat{\bf q}$ is defined by Eq.~(\ref{qcomponents}).  Combining these last
two equations with the results of Sec.~4 gives the asymptotic expression for
$\Phi^{(2)}$ shown in Eq.~(\ref{phi2asym}). 

To obtain an expression for the magnetic field we need the derivatives of $L$ and
$M$.  Making use of the fact that these are both diagonal matrices, we find that
\begin{equation}
    L^{-1/2} {\bf \nabla} L^{1/2} = -{1\over 2} L{\bf \nabla} L^{-1} 
       = -{1\over 2}h {\bf V} \end{equation}
and
\begin{equation}
    M^{-1/2} {\bf \nabla} M^{1/2} = -{1\over 2} M {\bf \nabla} M^{-1} 
     = -{1\over 2}M {\bf V}
\end{equation}
where $h=y_L +y_R +R +2b$.  In addition, we note that 
\begin{equation}
   L = 1 - h M^{-1}
\end{equation}
Using the last three identities, we can decompose the magnetic field ${\bf
B}^{(2)}$ as 
\begin{eqnarray}
   B_a^{(2)} &=& \epsilon_{abc} \left[ \partial_b A^{(2)}_c 
      + i  A^{(2)}_b  A^{(2)}_c \right]  \cr
    &=& L^{1/2}\left[ M^{1/2} b_a^{(A)} M^{1/2} + h M^{1/2} b_a^{(B)} M^{1/2}
      \right] L^{1/2} 
\end{eqnarray}
where
\begin{eqnarray}
    b_a^{(A)}  &=& 
   \epsilon_{abc} \left[ -{1\over 2}\partial_b ({\bf V} \times \bsigma)_c 
      -{1\over 2} \{M^{-1/2} \partial_b M^{1/2}, ({\bf V} \times \bsigma)_c \}
     + {i\over 4} ({\bf V} \times \bsigma)_b M ({\bf V} \times \bsigma)_c  
    \right] \cr
     &=& {1\over 2} \left( \sigma_a \partial_b V_b
            - \sigma_b \partial_b V_a  \right)
   +{1\over 4} V_b \{M, (V_a \sigma_b -V_b \sigma_a) \}
     + {i\over 4} ({\bf V} \times \bsigma)_b M V_a \sigma_b  
\end{eqnarray}
and
\begin{eqnarray}
   b_a^{(B)} &=& \epsilon_{abc} \left[ {1\over 2} V_b ({\bf V} \times
\bsigma)_c  
     -{i\over 4} ({\bf V} \times \bsigma)_b  ({\bf V} \times \bsigma)_c  
         \right]\cr
   &=&  V_a {\bf V}\cdot \bsigma - {\bf V}^2 \sigma_a
\end{eqnarray}
The last equation, together with Eqs.~(\ref{MsigmaM}-\ref{MyrM}),
immediately gives the terms in Eq.~(\ref{b2asym}) proportional to $h$.
To obtain the $h$-independent terms, it is useful to write
\begin{equation}
   M = {y_L + y_R \over y_L y_R {\bf V}^2} - {R\sigma_3 \over (y_L + y_R)^2 -R^2 }
      \equiv {y_L + y_R \over y_L y_R {\bf V}^2} - g \sigma_3
\end{equation}
One then finds that 
\begin{equation}  
 b_a^{(A)} = { ({\hat y}_L)_a  \over 2y_L} {\hat{\bf y}}_L \cdot \bsigma 
      + { ({\hat y}_R)_a  \over 2y_R} {\hat{\bf y}}_R \cdot \bsigma
      + g (V_a V_3 - {1\over 2}{\bf V}^2 \delta_{a3})
\end{equation}
which leads to 
\begin{equation}
   M^{1/2} b_a^{(A)}  M^{1/2} = { ({\hat y}_L)_a  \over 2y_L^2} 
     \left({\hat{\bf q}\cdot\bsigma + 1 \over 2 }\right) 
    + { ({\hat y}_R)_a  \over 2y_R^2} 
     \left({\hat{\bf q}\cdot\bsigma - 1 \over 2 }\right) 
     +{1\over 2} M\left[ { ({\hat y}_R)_a  \over y_R} 
                       - { ({\hat y}_L)_a  \over y_L}
            + g (2V_a V_3 - {\bf V}^2 \delta_{a3}) \right]
\end{equation}
After some algebra, one finds that the quantity in square brackets multiplying $M$
vanishes, thus leading to the final expression for ${\bf B}^{(2)}$, 
Eq.~(\ref{b2asym}).

\end{document}